\documentclass[aps,preprint,showpacs]{revtex4}
\usepackage{amssymb,epsfig}
\begin{document}

%\draft

\title {On the $SU(3)$ Symmetry-Breaking Corrections to Meson Distribution 
Amplitudes }

\author{V.~M.~Braun and A.~Lenz \\}

\affiliation{Institut f{\"u}r Theoretische Physik, Universit{\"a}t
          Regensburg, D-93040 Regensburg, Germany }

\date{\today}

\begin{abstract}
We consider constraints on the 
momentum fraction of the $K$ and $K^*$ meson  
carried by the strange quark that follow from exact operator identities,
similar to those for the divergence of the quark part of the 
QCD energy-momentum tensor. 
The existing QCD sum rule estimates are reanalyzed in this context. 
Our conclusions essentially support the constituent quark-model picture where 
the momentum fraction is roughly proportional to the constituent quark mass,
but the asymmetry turns out to be smaller compared to the naive quark
model estimates.
As a byproduct of this study,
we calculate the $SU(3)$-breaking quark-antiquark-gluon matrix 
elements that determine the leading conformal spin contributions to the 
asymmetry in twist-four distribution amplitudes of strange mesons
$K$ and $K^\ast$, and also update the estimate for the $SU(3)$ breaking 
for the quark-antiquark-gluon vacuum condensate.
\end{abstract}


%\tighten

%\narrowtext

\maketitle

%{\tableofcontents}

\section{Introduction}
\setcounter{equation}{0}

The $SU(3)$ flavor symmetry breaking effects in light-cone distribution amplitudes 
of light strange  mesons are attracting considerable interest in the 
context of the 
QCD description of exclusive $B$-meson decays. In particular, such effects play an important role 
in the extraction of the CKM angle $\gamma$ from $B \to \pi \pi, \pi K, K K$  decays \cite{Fleischer02}
in the framework of QCD factorization, see e.g. \cite{BBNS01}, and in light-cone sum 
rules for the semileptonic and rare radiative B-decay form factors \cite{massengrab1}. 
The question is far from being settled. 
%For example light-cone sum 
%rules for the semileptonic and rare radiative B-decay form factors \cite{massengrab1}  tend to predict 
%larger $SU(3)$ breaking compared to the corresponding lattice calculations \cite{massengrab2}.

The present work is motivated by the 
recent study \cite{BBog} which contains an update of the QCD sum rules for 
the lowest moments of the distribution amplitudes (DA) of $K$ and $K^*$ mesons.
Taking at face value, the sum rules derived in \cite{BBog} suggest that the 
strange quark carries a {\it smaller} momentum fraction of the meson compared to
that of the nonstrange (anti)quark, which is unexpected and contradicts the intuition 
inherited from quark models. Inspection of the sum rules in \cite{BBog} reveals that they 
suffer from considerable cancellations so it is not clear whether their accuracy 
is sufficient to warrant this conclusion. In this paper we 
suggest a complementary approach that is based on the use of operator identities
that are reminiscent of those for the divergence of the quark part of the energy-momentum tensor.

The general idea of using such identities to get rid of cancellations of 
leading contributions in the sum rules  belongs to A.~Kolesnichenko who 
employed the same technique to calculate the momentum fraction of the nucleon carried by gluons \cite{Kol84}.
A similar approach was advocated by M.~Neubert in \cite{Neubert93}, albeit in a different context.  
In the present case, the rationale is to understand the physical reasons for 
the cancellations observed in \cite{BBog} and derive new sum rules where the  
cancellations are less pronounced.      
Our conclusions essentially 
support the constituent quark-model picture where the quark and the antiquark have
equal velocities and hence the momentum fraction is roughly proportional to 
the constituent quark mass. Numerically, however, the effect is smaller. 
As a byproduct of this study, we calculate the matrix 
elements of quark-antiquark-gluon operators that determine the $SU(3)$-breaking  
contributions in twist-four DAs of strange mesons, and also 
reanalyze the existing estimates of the $SU(3)$ breaking in the mixed 
quark-gluon vacuum condensate. 

\section{Longitudinally polarized $K^*$ meson}

As a prime example, we consider the leading-twist DA of the longitudinally polarized
$K^*$ meson defined as \cite{CZreport,BB96,BBKT98,BB99,BBog}
\begin{equation}
 \langle 0| \bar u(-\alpha n)\not\!n\, s(\alpha n)|K^{\ast}(q,\lambda)\rangle =
 (e^{(\lambda)}\cdot n)f_{K^*}m_{K^*}
 \int\limits_0^1du\,e^{-iqn\alpha(2u-1)}\phi_{K^*}(u,\mu^2)\,.
\label{DAKlong}
\end{equation}
Here $q_\mu$ is the meson momentum, $q^2 = m_{K^*}^2$, $e^{(\lambda)}_\mu$ is the polarization vector, $n_\mu$ is a light-like vector, $n^2=0$, $\alpha$ is an arbitrary real number and a path-ordered gluon exponential is understood 
between the quark fields on the l.h.s. 
The variable $u$ is the momentum fraction carried by the 
strange quark and $\mu$ is  the factorization scale. 
Breaking of the $SU(3)$ flavor symmetry leads to a nonvanishing first moment 
of the DA
\begin{equation}
 \langle 0| \bar u \not\!n\, i\stackrel{\leftrightarrow}{D}\!\cdot n\, s |K^{\ast}(q,\lambda)\rangle =
 (e^{(\lambda)}\cdot n)(q\cdot n)f_{K^*}m_{K^*} \phi_1^{K^*}
\label{phi1Klong}
\end{equation}
where $D_\mu = \partial_\mu -ig A_\mu$, $\stackrel{\leftrightarrow}{D} = \stackrel{\to}{D}-
\stackrel{\gets}{D}$ is the covariant derivative and 
we introduced the notation $\phi_1^{K^*}$ for the integral
\begin{equation}
 \phi_1^{K^*}(\mu)  = \int\limits_0^1du\,(2u-1)\phi_{K^*}(u,\mu^2) = \langle u_s-u_u\rangle\,.
\end{equation}
The value of $\phi_1^{K^*}$ provides a measure of an overall asymmetry of the quark momentum fraction 
distribution. In particular  $\phi_1^{K^*} > 0 $ corresponds to a 
larger momentum fraction of $K^*$ carried 
by the strange quark, and vice versa. It is related to the often used first Gegenbauer moment
of the DA (see e.g. \cite{BBog}) as
\begin{equation}
         a^{K^*}_1 = \frac53 \phi_1^{K^*}\,.
\end{equation}

As follows from 
(\ref{phi1Klong}) 
the first moment of the DA is given by the meson-to-vacuum matrix
element of the flavor-nonsinglet analogue of the quark part of the QCD energy momentum tensor. We define
\begin{equation}
  O_{\mu\nu} = \frac12 \bar u \gamma_\mu i\stackrel{\leftrightarrow}{D}_\nu s +
               \frac12 \bar u \gamma_\nu i\stackrel{\leftrightarrow}{D}_\mu s -
               \frac14 g_{\mu\nu} \bar u\, i\!\not{\!\!\stackrel{\leftrightarrow}{D}}\, s 
\label{O}
\end{equation}  
so that 
\begin{equation}
   \langle 0|O_{\mu\nu}|K^{\ast}(q,\lambda)\rangle =
 \frac12 f_{K^*}m_{K^*}\phi_1^{K^*} (e^{(\lambda)}_\mu q_\nu+e^{(\lambda)}_\nu q_\mu)\,.
\end{equation}
In this paper we analyse the consequences of an exact operator identity that relates  the divergence 
of $O_{\mu\nu}$ with a quark-antiquark-gluon operator
\begin{equation}
   \partial^\mu O_{\mu\nu} = 2i\, \bar u igG_{\nu\mu}\gamma^\mu s -
         \frac{i}{2}(m_s-m_u)\partial^\mu \bar u \sigma_{\mu\nu}s
         -i (m_s^2-m_u^2)\bar u\gamma_\nu s - \frac14 (m_u+m_s)\partial_\nu \bar u s 
\label{Id:vector}
\end{equation}
where we omitted terms that vanish by virtue of QCD equations of motion. The derivation of (\ref{Id:vector})
is straightforward. 
Note that the quark-antiquark-gluon operator gets mixed with the other operators on the r.h.s. 
of (\ref{Id:vector}) upon renormalization. This mixing is triangular and such that the sum of terms on the 
r.h.s. of (\ref{Id:vector}) is renormalized multiplicatively with the anomalous dimension 
of $O_{\mu\nu}$. 
Taking the matrix element between the vacuum and the $K^*$-meson state and using 
standard notations 
\begin{eqnarray}
   \langle 0| \bar u \gamma_\nu  s|K^{\ast}(q,\lambda)\rangle &=&  e^{(\lambda)}_\nu f_{K^*}m_{K^*}\,,
\nonumber\\
   \langle 0| \bar u \sigma_{\mu\nu}s|K^{\ast}(q,\lambda)\rangle &=& 
 i(e^{(\lambda)}_\mu q_\nu - e^{(\lambda)}_\nu q_\mu)f^\perp_{K^*}
\label{fKstar}
\end{eqnarray}
we obtain an exact relation  
\begin{equation}
 \phi_1^{K^*} = - 4 \kappa_4^{K^*} - \frac{m_s-m_u}{m_{K^*}} \frac{f^\perp_{K^*}}{f_{K^*}} 
     + 2 \frac{m_s^2-m_u^2}{m^2_{K^*}}
\label{relation1}
\end{equation}
where we have introduced a new coupling $\kappa_4^{K^*}$ by the matrix element
\begin{equation}
  \langle 0| \bar u igG_{\nu\mu}\gamma^\mu s |K^{\ast}(q,\lambda)\rangle =  e^{(\lambda)}_\nu f_{K^*}m^3_{K^*}
 \kappa_4^{K^*}\,. 
\label{kap4star}
\end{equation}
The last term on the r.h.s. of (\ref{Id:vector}) does not contribute. 
Note that for massless quarks the quark-gluon operator in (\ref{kap4star}) has exotic quantum 
numbers $J^{\rm PC}=1^{-+}$ and indeed it was used for studies of quark-antiquark-gluon states 
in Refs.~\cite{BDY82}. It follows that the matrix element in (\ref{kap4star}) is in general 
first order in the $SU(3)$ breaking, $\kappa_4^{K^*} = {\mathcal O}(m_s)$.

Since the quark masses and the couplings in (\ref{fKstar}) are relatively well known, the identity in 
(\ref{relation1}) establishes a relation between the $SU(3)$ breaking in momentum fractions carried 
by the quark and the antiquark in the $K^*$ meson and the quark-antiquark-gluon Fock state admixture in the
wave function. In the remainder of this work we analyze the consequence of this relation and a similar 
identity for the DA of the pseudoscalar $K$ meson that we derive later on. In particular, we will
argue that using Eq.~(\ref{relation1}) one can write down more accurate QCD sum rules.

As the first step, however, let us discuss this relation in the context of a simple nonrelativistic 
quark model. The operator relation (\ref{Id:vector}) remains true with the substitution 
of QCD quark operators and current quark masses by the effective constituent quark fields 
$\bar u \to \bar U, s \to S$ and 
constituent quark masses $m_u \to M_U, m_s \to M_S$, respectively. The matrix element of 
the operator built of constituent quark fields and a gluon (quark model analogue of $\kappa_4^{K^*}$)
vanishes by assumption, and in the nonrelativistic limit also $f_{K^*}= f^\perp_{K^*}$. 
We further assume $M_S+M_U \simeq m_{K^*}$ and $M_S-M_U\simeq m_s-m_u \simeq m_s$.
Eq.~(\ref{relation1}) in this limit becomes
\begin{equation}
 \phi_1^{K^*} = -  \frac{M_s-M_u}{m_{K^*}} + 2 \frac{M_S^2-M_U^2}{m^2_{K^*}} = 
                        + \frac{M_s-M_u}{m_{K^*}} \sim \frac{m_s}{m_{K^*}} \sim 0.15 
\end{equation}
which is the expected result: the momentum fractions of the quark and the antiquark 
are proportional to their masses. Note that the term linear in $M_S$ gives the same 
number but with a negative sign, and the sign is corrected by the contribution of the 
last term $\sim M_S^2-M_U^2$ which in the quark model is first order in the $SU(3)$ breaking parameter
$(M_S-M_U)/(M_S+M_U)$. Going back to QCD the situation changes drastically. In this case
the third, last term in Eq.~(\ref{relation1}) is second order in the $SU(3)$ breaking and its 
role is overtaken by the quark-antiquark-gluon matrix element which, in this sense, corresponds 
to the contribution of gluons `hidden' inside the constituent quarks. Other way around, 
the role of gluon degrees of freedom is overtaken in the quark model by the constituent quark mass.        
{}From this equivalence we  obtain, in (the simplest version of) the quark model
\begin{equation}
   \kappa_4^{K^*}\Big|_{\rm QM} = -\frac12 \frac{m_s}{m_{K^*}} \sim -0.07 + {\cal O}(m^2_s/m^2_{K^*})
\end{equation}
The QCD sum rule calculation of this quantity gives instead 
\begin{equation}
   \kappa_4^{K^*}\Big|_{\rm SR} = -(0.050\pm 0.0010)   
\label{kappaKstar}
\end{equation}
at the scale 1 GeV, which is comparable. 
This calculation is rather technical and presented in detail in Appendix A where 
we also elaborate on the difference of this sum rule to the one considered in \cite{BBog}.
Combining this result with $m_s(1$~GeV$)\simeq 130$~MeV and the estimate $f^\perp_{K^*}/f_{K^*}\simeq 0.8$
supported by both QCD sum rules \cite{BB96} and lattice calculations \cite{lat:ratio1,lat:ratio2},  we obtain
(at 1 GeV)
\begin{equation}
   \phi_1^{K^*} = 0.06\pm 0.04\,, \qquad  a_1^{K^*} = 0.10\pm 0.07\,,  
\label{resultKpar}
\end{equation}
i.e. smaller than in quark model, but still of positive sign. 
In this estimate we have discarded 
the constribution of the last, third term in (\ref{relation1}) because it is  
${\cal O}(m^2_s)$ while the sum rules are written to the ${\cal O}(m_s)$ accuracy only.
Numerically this omitted contribution is $2m_s^2/m_{K^*}^2 = +0.043$ which is rather large in view 
of the small number in (\ref{resultKpar}) and, if added, brings our result 
closer to the quark model:
\begin{equation}
   \phi_1^{K^*} = 0.10\pm 0.05\,, \qquad  a_1^{K^*} = 0.17\pm 0.08\,.  
\label{result2Kpar}
\end{equation}

\section{Pseudoscalar $K$ meson}

The same method can be used for pseudoscalar mesons. The leading-twist $K$ meson DA  is defined as  
\begin{equation}
 \langle 0| \bar u(-\alpha n)\not\!n\, \gamma_5 s(\alpha n)|K(q)\rangle =
  i f_{K} (q\cdot n) \int\limits_0^1du\,e^{-iqn\alpha(2u-1)}\phi_K(u,\mu^2)\,,
\label{DAK}
\end{equation}
and its first moment which quantifies the asymmetry of $\phi_K(u,\mu^2)$ is given by 
the matrix element 
\begin{equation}
 \langle 0| \bar u \not\!n\gamma_5  i\!\stackrel{\leftrightarrow}{D}\!\cdot n\, s |K(q)\rangle =
 i f_K (q\cdot n)^2\phi^K_1\,,\qquad 
 \phi^K_1(\mu)  = \int\limits_0^1du\,(2u-1)\phi_K(u,\mu^2)\,.
\label{phi1}
\end{equation}
{}Following the same procedure as above we introduce the symmetric and traceless operator
\begin{equation}
  O^5_{\mu\nu} = \frac12 \bar u \gamma_\mu\gamma_5\, i\!\stackrel{\leftrightarrow}{D}_\nu s +
               \frac12 \bar u \gamma_\nu\gamma_5\, i\!\stackrel{\leftrightarrow}{D}_\mu s -
               \frac14 g_{\mu\nu} \bar u\, i\!\not{\!\!\stackrel{\leftrightarrow}{D}}\gamma_5\, s 
\label{O5}
\end{equation}  
so that 
\begin{equation}
   \langle 0|O^5_{\mu\nu}|K(q)\rangle = i f_K \phi_1^K\left[
    q_\mu q_\nu - \frac14 m_K^2 g_{\mu\nu}\right].
\end{equation}
The divergence of $O^5_{\mu\nu}$ is easily calculated to be 
\begin{equation}
   \partial^\mu O^5_{\mu\nu} = 2i\, \bar u igG_{\nu\mu}\gamma^\mu \gamma_5 s +
         \frac{i}{2}(m_s+m_u)\partial^\mu \bar u \sigma_{\mu\nu}\gamma_5s
         -i (m_s^2-m_u^2)\bar u\gamma_\nu\gamma_5 s - \frac14 (m_u-m_s)\partial_\nu \bar u\gamma_5 s\,. 
\label{Id:pseudo}
\end{equation}
Using the standard defintions
\begin{equation}
   \langle 0| \bar u \gamma_\nu\gamma_5  s|K(q)\rangle =  i f_{K} q_\nu\,,\qquad
   \langle 0| \bar u i\gamma_5 s|K(q)\rangle = \frac{f_K m_K^2}{m_u+m_s}\,,
\label{fK}
\end{equation}
and introducing the notation for the quark-antiquark-gluon coupling
\begin{equation}
  \langle 0| \bar u igG_{\nu\mu}\gamma^\mu\gamma_5  s |K(q)\rangle = 
  i f_{K} q_\nu m^2_{K} \kappa_4^{K}\,, 
\label{qGq}
\end{equation}
we obtain the identity
\begin{equation}
  \frac32\phi^K_1 = - 4\kappa_4^{K} + 2 \frac{m_s^2-m_u^2}{m_K^2}-\frac12 \frac{m_s-m_u}{m_s+m_u}\,.
\label{relation2}
\end{equation}
Note that the operator with the $\sigma$-matrix on the r.h.s. of (\ref{Id:pseudo}) does not contribute.

Although this relation looks similar to the one for vector mesons (\ref{relation1}), the counting 
of the $SU(3)$-breaking is different since $m_K^2\sim {\cal O}(m_s)$. By inspection one finds 
that in the chiral limit of Eq.~(\ref{relation2}) the first and the third term on the r.h.s. 
are of order one, while the remaining contributions are ${\cal O}(m_s)$. 
It follows that the two large terms have to cancel against each other
so that the quark-antiquark-gluon matrix element can be calculated exactly in this limit
\begin{equation}
  \kappa_4^K = -\frac18 +\mbox{\cal O}(m_s)\,.
\label{surprise}
\end{equation}
The expression in (\ref{surprise}) presents one of the main results of this paper.  

The cancellation between the first and the third contribution on the r.h.s. of Eq.~(\ref{relation2})
in the chiral limit implies that this relation is less useful 
for the stated purpose of the determination of $\phi^K_1$ since to this end the 
quark-antiquark-gluon matrix element in (\ref{qGq}) has to be calculated to second order 
in the $SU(3)$ breaking, alias first order for $\kappa_4^{K}$. On the other hand, 
the result in (\ref{surprise}) is of considerable interest by itself. It turns out that 
the existing estimates of chiral symmetry breaking in the quark-gluon condensate
\begin{equation}
          \delta_5 = 1-\frac{\langle \bar s \sigma gGs\rangle}{\langle \bar u \sigma gGu\rangle}
\label{delta5}
\end{equation}
largely rely on the comparison of different QCD sum rule predictions for 
this coupling \cite{KKZ87,BD92,AM95}, and knowledge of the exact value in the chiral limit 
provides an important consistency check. 
The corresponding sum rules are considered in Appendix B. When combined with similar sum rules 
for the vector channel (Appendix A), we obtain an updated estimate for this important parameter
\begin{equation}
     \delta_5 = 0.15\pm 0.1
\label{d5final}
\end{equation}
which is of general interest to QCD sum rule practitioneers. With a partial taking into account 
of higher-order effects in the $SU(3)$ breaking that are encoded in the particle spectra and 
physical (measured) values of the decay constants, we further obtain
\begin{equation}
    \kappa_4^{K}\Big|_{\rm SR} = -(0.11\pm 0.03)\,.   
\label{kaSR}
\end{equation}
This number coincides within errors with the result in the $SU(3)$ limit and leads to
the value of the first moment
\begin{equation}
   \phi^K_1 = 0.06\pm 0.07\,, \qquad    a^K_1 = 0.10\pm 0.12\,,  
\label{resultK}
\end{equation}
which is positive but small and has a large uncertainty. 

\section{Conclusions}

In this paper we have analysed consequences of exact operator identities that 
relate the difference in momentum fraction carried by strange and nonstrange quarks 
in $K$ and $K^*$ mesons to matrix elements of certain quark-antiquark-gluon operators. 
Our results essentially support the quark model picture in which 
heavier consituents carry a larger momentum fraction, but also suggest that this pattern 
may change when going over from heavy to light mesons. It would be very interesting to check 
the quark mass dependence using lattice calculations.
The $SU(3)$ breaking quark gluon matrix elements that are estimated in this work are interesting 
in their own right and, in particular, they define $SU(3)$ breaking contributions to twist-four 
meson DAs \cite{BBKT98,BB99}. A detailed study of $SU(3)$ breaking in higher-twist
DAs goes beyond the tasks of this work. We note, however, that in several cases
such contributions have {\it lower}\ conformal spin compared to $SU(3)$-symmetric contributions
and correspond to the ``true'' asymptotic distributions at large scales. 
They can have significant effect on the $SU(3)$ breaking in heavy meson decay form factors.  

The numerical estimates reported in this work have been obtained by the analysis of four 
different QCD sum rules for the correlation functions of relevant quark-gluon operators 
with suitable currents. It is important to stress that all four sum rules are consistent with one another, 
and we also explained why the sum rule derived in \cite{BBog} cannot be used.
This sum rule presents an example of the so-called ``nondiagonal'' sum rules, with different 
chiral structure of the two participating currents, and it is claimed sometimes that these are notoriously unreliable.
%The distinction between  ``useful'' and ``useless'' sum rules is sometimes formulated 
%as the dilemma of using ``diagonal'' or ``nondiagonal'' sum rules, with either the same or the different 
%chiral structure of the currents. 
We do not think that such a general conclusion is warranted; rather one has to analyze
the sum rules case by case, and conclusions can vary.

\section*{Note added}
When this paper was in writing, the work \cite{KMM04} appeared where the authors consider 
sum rules for the $K$-meson DA and arrive to similar conclusions. Our method 
is, however, different so that the two studies complement each other.

\section{Appendix A}
{}For the discussion  of the difference of our results with the sum rules 
considered in \cite{BBog} we need the following three correlation functions:
\begin{equation}
 i\int d^4y e^{-iqy}\langle 0|{\rm T}\{O_{\mu\nu}(0) \bar s(y)\sigma_{\alpha\beta}u(y)\}|0\rangle=
     i[q_\mu q_\alpha g_{\beta\nu}+q_\nu q_\alpha g_{\beta\mu} -
      q_\mu q_\beta g_{\alpha\nu}-q_\nu q_\beta g_{\alpha\mu}]W(q^2)\,,
\label{W}
\end{equation}
\begin{equation}
 i\int d^4y e^{-iqy}\langle 0|{\rm T}\{\partial_\mu(\bar u(0)\sigma_{\mu n}s(0))
    s(y)\sigma_{\alpha n }u(y)\}|0\rangle = -i (qn)n_\alpha \Pi(q^2)\,,
\label{P}
\end{equation}
where we used a shorthand notation $\sigma_{\mu n} \equiv \sigma_{\mu\nu} n^\nu$ etc., 
and 
\begin{equation}
 i\int d^4y e^{-iqy}\langle 0|{\rm T}\{\bar u(0) igG_{\mu\nu}{\gamma^\nu}s(0) 
  \bar s(y)\sigma_{\alpha\beta}u(y)\}|0\rangle= i[q_\beta g_{\mu\alpha}-q_\alpha g_{\mu\beta}]\Pi_G^{(\sigma)}(q^2)\,.
\label{PG}
\end{equation}
The operator identity in Eq.~(\ref{Id:vector}) implies the relation between the correlation functions
\begin{equation}
  2 q^2 W(q^2) = 4 \Pi_G^{(\sigma)}(q^2) +m_s\Pi(q^2) +~\mbox{\rm contact terms}~+{\cal O}(m_s^2) 
\label{recor}
\end{equation} 
which provides one with a nontrivial check if all the three functions are computed independently.  
The contact terms $\sim \langle \bar q q\rangle$ can easily be calculated but are of no interest 
for what follows since they are eliminated by the subsequent Borel transformation. The operator product expansion 
(OPE) for $W(q^2)$ reads
\begin{eqnarray}
  2 W(q^2) &=& -\frac{m_s}{8 \pi^2}\ln\frac{-q^2}{\mu^2}
                \left[ 1 + \frac{\alpha_s}{3\pi}\left(11+\frac13 \ln\frac{-q^2}{\mu^2}\right)\right]
\nonumber\\
&&{}+\frac{\langle \bar u u\rangle -\langle \bar s s \rangle}{q^2}
               \left[-1+\frac{20}{9}\frac{\alpha_s}{\pi} \left(\frac53 - \ln\frac{-q^2}{\mu^2}\right)\right]
\nonumber\\
&&{}- \frac{m_s}{q^4}\left \langle\frac{\alpha_s}{\pi}G^2\right\rangle 
               \left(-\frac{19}{24} +\frac12\ln\frac{-q^2}{\mu^2}\right) 
    - \frac{\langle \bar u \sigma gGu\rangle -\langle\bar s\sigma gG s \rangle}{q^4}
\nonumber\\
&&{}+0\cdot \frac{1}{q^6} m_s\alpha_s\langle \bar q q \rangle^2+\ldots
%&&{}+\frac{16\pi}{81q^6} m_s\alpha_s\langle \bar q q \rangle^2+\ldots
\label{WOPE}
\end{eqnarray}
This expansion can be extracted from the expressions given in \cite{BBog} apart from the 
contribution of the gluon condensate and the four-quark condensate $\langle\bar q q \rangle^2$ which 
are new results. In the latter case the usual vacuum factorization approximation has been used.  
The leading-order quark condensate contribution in (\ref{WOPE}) corresponds to the contact 
term in (\ref{recor}).
Since the correlation function $\Pi(q^2)$ enters Eq.~(\ref{recor})  multiplied by $m_s$, we only 
need its OPE in the $SU(3)$ symmetry limit which is well known \cite{RRY84,BB96}
\begin{eqnarray}
 \Pi(q^2) &=& -\frac{1}{8\pi^2} q^2 \ln\frac{-q^2}{\mu^2} 
             \left[ 1 + \frac{\alpha_s}{3\pi}\left(\frac73 + \ln\frac{-q^2}{\mu^2}\right)\right]
\nonumber\\&&{}
             -\frac{1}{24q^2} \left \langle\frac{\alpha_s}{\pi}G^2\right\rangle +
             \frac{208\pi}{81q^4}\alpha_s\langle \bar q q \rangle^2+\ldots 
\label{POPE}
\end{eqnarray}  
Finally, we obtain by explicit calculation
\begin{eqnarray}
   \Pi_G^{(\sigma)}(q^2) &=&  -\frac{m_s\alpha_s}{144\pi^3} q^2 \ln\frac{-q^2}{\mu^2} 
       \left( 13 - \ln\frac{-q^2}{\mu^2}\right)
\nonumber\\
&&{} -\frac{5}{9}\frac{\alpha_s}{\pi}(\langle \bar u u\rangle -\langle \bar s s \rangle) \ln\frac{-q^2}{\mu^2}
\nonumber\\
&&{} -\frac{1}{4}\frac{\langle \bar u\sigma gG  u\rangle -\langle\bar s\sigma gG  s \rangle}{q^2} 
     +\frac{m_s}{8 q^2}\left \langle\frac{\alpha_s}{\pi}G^2\right\rangle
       \left[ \frac53 -  \ln\frac{-q^2}{\mu^2}\right]
\nonumber\\
&&{} -\frac{52\pi}{81q^4}m_s\alpha_s\langle \bar q q \rangle^2+\ldots   
%&&{} -\frac{16\pi}{27q^4}m_s\alpha_s\langle \bar q q \rangle^2+\ldots   
\label{PGOPE}
\end{eqnarray}
In all cases $\mu$ is the $\overline{\rm MS}$ normalization scale and $\alpha_s=\alpha_s(\mu)$.
Note that the scale dependence of the gluon condensate contribution corresponds to an infrared divergence
which occurs due to the mixing of the operators $\bar s\sigma gG  s$ and $m_s G^2$.
We have checked that the identity in Eq.~(\ref{recor}) is satisfied for perturbation theory, quark condensate 
and the mixed quark gluon condensate contributions, and we used this relation 
to calculate the contributions of the gluon condensate and the four-quark condensate to $W(q^2)$, see 
Eq.~(\ref{WOPE}). 

In addition, we have calculated the correlation function of the relevant quark-antiquark-gluon
operator with the vector current  
\begin{equation}
 i\int d^4y e^{-iqy}\langle 0|{\rm T}\{\bar u(0) igG_{\mu\nu}{\gamma^\nu}s(0) 
  \bar s(y)\gamma_{\alpha}u(y)\}|0\rangle= 
  g_{\mu\alpha}\Pi_G^{(v)}(q^2)+{\mathcal O}(q_\mu q_\alpha)\,,
\label{PGvec}
\end{equation}
\begin{eqnarray}
 \Pi_G^{(v)}(q^2) &=&  \frac{m_s\alpha_s}{3\pi}
   \left[ \frac53 \langle\bar u u\rangle + \langle\bar s s\rangle\right]\ln\frac{-q^2}{\mu^2}
+\frac{m_s}{4q^2} \langle \bar u\sigma gG  u\rangle
  + \frac{m_s}{12q^2}\langle\bar s\sigma gG  s \rangle
\nonumber\\&&{}
-\frac{8\pi\alpha_s}{27 q^2}\big[\langle \bar u u\rangle^2 -\langle \bar s s \rangle^2\big].
\label{PGvecOPE}
\end{eqnarray}
In this case contributions of the perturbation theory and of the gluon condensate
are of order ${\mathcal O}(m_s^2)$ and they were omitted. 

The sum rules are constructed in a usual way, equating the OPE to the contribution of the lowest 
intermediate state ($K^*$-meson) for the intermediate region of $q^2$, making the Borel transformation 
and substracting the continuum contribution above a certain threshold $s_0$. 
We obtain
\begin{eqnarray}
 f_{K^*}f^\perp_{K^*}m_{K^*}\phi_1^{K^*} e^{-m^2_{K^*}/M^2} &=&
   -\frac{m_s}{8 \pi^2}\int_0^{s_0} ds\, e^{-s/M^2}
                \left[ 1 + \frac{\alpha_s}{3\pi}\left(11+\frac23 \ln\frac{s}{\mu^2}\right)\right]
\nonumber\\
&&\hspace*{-0.2cm}{}- (\langle \bar u u\rangle -\langle \bar s s \rangle)\left[
       1-\frac{20}{9}\frac{\alpha_s}{\pi}\left(\frac53+\gamma_E-\ln\frac{M^2}{\mu^2}+
         \!\int_{s_0}^\infty\! \frac{ds}{s}\,e^{-s/M^2}\right)\right]
\nonumber\\
&&{}+ \frac{\langle \bar u\sigma gG  u\rangle -\langle\bar s\sigma gG  s \rangle}{M^2} 
\nonumber\\
&&{}+ \frac{m_s}{2 M^2}\left \langle\frac{\alpha_s}{\pi}G^2\right\rangle\left[
      -\frac{7}{12}-\gamma_E+\ln\frac{M^2}{\mu^2}+
         M^2\! \int_{s_0}^\infty\! \frac{ds}{s^2}\,e^{-s/M^2}\right],
%\nonumber\\
%&&{}+\frac{8\pi}{81 M^4} m_s \alpha_s\langle \bar q q\rangle^2\,,               
\label{Patricia}
\end{eqnarray}     
\begin{eqnarray}
\hspace*{-2cm}
  f_{K^*}f^\perp_{K^*}m_{K^*}^3(-4\kappa_4^{K^*}) e^{-m^2_{K^*}/M^2} &=&
  -\frac{m_s\alpha_s}{9\pi^3}\int_0^{s_0} ds\,s\, e^{-s/M^2}
  \left[\frac{13}{4}-\frac{1}{2} \ln\frac{s}{\mu^2}\right]
\nonumber\\
&&{}-\frac{20}{9}\frac{\alpha_s}{\pi} (\langle \bar u u\rangle -\langle \bar s s \rangle)
    \int_0^{s_0} ds\, e^{-s/M^2}
\nonumber\\
&&{} - (\langle \bar u\sigma gG  u\rangle -\langle\bar s\sigma gG  s \rangle)
\nonumber\\
&&{}- \frac{m_s}{2} \left \langle\frac{\alpha_s}{\pi}G^2\right\rangle\left[
       -\frac53-\gamma_E + \ln\frac{M^2}{\mu^2}
        -\int_{s_0}^\infty \frac{ds}{s}\,e^{-s/M^2}
         \right]
\nonumber\\
&&{}+\frac{208\pi}{81M^2}m_s\alpha_s \langle \bar q q\rangle^2\,,
%&&+\frac{64\pi}{27M^2}m_s\alpha_s \langle \bar q q\rangle^2\,,
\label{we1}
\end{eqnarray}
\begin{eqnarray}
\hspace*{-2.5cm}
  f^2_{K^*}m_{K^*}^4(-4\kappa_4^{K^*}) e^{-m^2_{K^*}/M^2} &=&
 -\frac{32\alpha_s}{9\pi}m_s\langle\bar q q\rangle 
  \int_0^{s_0}ds\, e^{-s/M^2}
 -\frac43 m_s \langle \bar q\sigma gG  q\rangle
\nonumber\\&&{} 
 +\frac{32\pi\alpha_s}{27} \big[\langle \bar u u\rangle^2 -\langle \bar s s \rangle^2\big],
\label{we2}
\end{eqnarray}
from the correlation functions $W(q^2)$, $\Pi_G^{(\sigma)}(q^2)$ and $\Pi_G^{(v)}(q^2)$  
respectively. $M^2$ is the Borel parameter. We use the notation 
$\langle\bar q q\rangle$ and $\langle \bar q\sigma gG  q\rangle$ for  
the condensates in case that distinguishing between strange and non-strange quarks 
is beyond our accuracy.

First, let us discuss the general structure of the
sum rules in some detail. The first of them, Eq.~(\ref{Patricia}), is essentially the 
sum rule considered in Ref.~\cite{BBog}, apart from a new contribution of the gluon 
condensate which is numerically not very significant. 
The two leading-order contributions 
to this sum rule, the perturbation theory and the quark condensate, have opposite sign and 
tend to cancel each other almost exactly. It is this unpleasant cancellation that motivated 
the present study and indeed we see that the both contributions are absent in the sum rule (\ref{we1})
in which case all terms include the strong coupling. The use of the operator identity in 
(\ref{Id:vector}) allows to include these contributions, effectively, in the second term on the 
r.h.s. of the relation (\ref{relation1}). Inspection of Eq.~(\ref{we1}) shows that 
in this sum rule all terms have the same positive sign with the only exception for the $\alpha_s$ perturbative 
correction, which is negative. Since there are no large cancellations, we expect the sum rule 
(\ref{we1}) is  more reliable. 
In addition, we observe that the perturbative correction $\sim {\mathcal O}(\alpha_s)$
to the quark condensate contribution in the sum rule (\ref{Patricia}) amounts to nearly 80\% 
of the leading-order contribution. Since the similar correction to mixed quark-gluon condensate 
has not been calculated, this results in an artificial, as we think,  suppression of 
the (positive) quark condensate contribution compared to the (negative) contribution
of the mixed condensate. We believe that in such a situation neglecting the  
$\sim {\mathcal O}(\alpha_s)$ corrections to the condensates altogether  would be a better option.    
As far as the sum rule in (\ref{we1}) is concerned, its biggest drawback is probably the strong 
sensitivity to the $SU(3)$ symmetry breaking in mixed quark-gluon condensate, which is not 
well known. Hence this sum rule can be reliable but probably has poor accuracy.
{}For this reason we also consider an alternative sum rule for the same quantity, Eq.~(\ref{we2}),  
in which case the structure of the OPE is very different. Comparing the sum rules in 
(\ref{we1}) and (\ref{we2}) we can judge upon the consistency of our approach.

After these preliminary remarks, we proceed to the numerical analysis and have to specify 
values of the sum rule parameters as the first step. 
The status of the strange quark mass determinations is at present rather controversial. While QCD sum 
rules and older lattice calculations obtain values in the range $m_s(2$~ GeV$)\sim 100-110$~MeV,
see e.g. \cite{JOP02}, some new lattice studies \cite{HPQCD} favor much smaller values
of order  $m_s(2$~GeV$)\sim 80$~MeV. For the numerical analysis in this work we adopt
\begin{equation}
                    m_s(1\,\mbox{\rm GeV}) = 130\pm 20~\mbox{\rm MeV} 
\end{equation} 
which corresponds to $m_s(2$\,GeV$) = 100\pm 15$~MeV.
For the other entries we use $\alpha_s(1$~GeV$)=0.5$, 
  $f_{K^*}= 220$~MeV, $f^\perp_{K^*}/f_{K^*}=0.8$ \cite{lat:ratio1,lat:ratio2}, $s_0^{K^*}=1.8$~GeV$^2$
and the vacuum condensates $\langle \bar u u\rangle = -(240$~MeV)$^3$,
$ \left \langle\frac{\alpha_s}{\pi}G^2\right\rangle = 0.012$~GeV$^4$, 
$\langle \bar u\sigma gG  u\rangle  = m_0^2 \langle \bar u u\rangle$ with $m_0^2=0.8$~GeV$^2$,
$\langle \bar s s\rangle = 0.8 \langle \bar u u\rangle$, and 
$\langle \bar s\sigma gG  s \rangle  = (1-\delta_5) \langle \bar u\sigma gG  u\rangle$
where we vary $\delta_5$ in the range $\delta_5 =0.2\pm 0.2$. 

\begin{figure}[t]
\centerline{\epsfysize5.7cm\epsffile{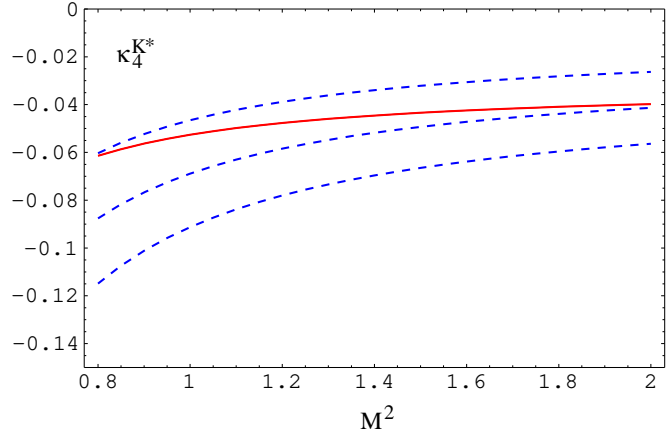}}
\caption[]{\small
 The quark-antiquark-gluon mattrix element  $\kappa_4^{K^*}$ from the sum rules 
(\ref{we2}) (solid red curve) and  (\ref{we1}) (dashed blue curves) as a function of the Borel parameter.
The three dashed curves are obtained using $\delta_5=0$, 0.1 and 0.2 from top to bottom, respectively. 
 }
\label{kappa}
\end{figure}

The results for the quark-antiquark-gluon matrix element  $\kappa_4^{K^*}$ from the sum rules 
(\ref{we1}) and  (\ref{we2}) as a function of the Borel parameter are shown in Fig.~\ref{kappa}.
The solid curve corresponds to the sum rule in (\ref{we1}), whereas
the three dashed curves (from top to bottom) are obtained from the sum rule (\ref{we1}) 
using $\delta_5 =0$, 0.1 and 0.2, respectively.  
We observe a reasonable agreement between the two sum rules,
 which is non-trivial since they  are very different.  
The sum rule in (\ref{we2}) appears to be significantly more stable than that in 
(\ref{we1}) and also less sensitive to variations of the sum rule parameters. We, therefore, 
use this sum rule for the final estimate of $\kappa^{K^*}_4$ quoted in (\ref{kappaKstar}).
Imposing the requirement that the two sum rules agree identically, we can determine the 
value of the $SU(3)$ breaking parameter $\delta_5 \simeq 0.05$, which is within the commonly accepted 
range. We will return to the discussion of this parameter in Appendix B. 

One may try to 
improve the stability of the sum rules  by adding the contribution of a second resonance on the 
phenomenological side, with the mass of order 1300--1400 MeV. 
In this way, both sum rules can be made perfectly stable 
and the estimate for the absolute value of $\kappa^{K^*}_4$ gets increased by ca.\ 30--50\%.
This increase, however, can well be an artifact of neglecting higher power corrections in the OPE. 
In particular, one should not be mislead by the smallness of the  $1/M^2$ correction to (\ref{we1})
that is due to the contribution of four-fermion operators. This correction is very likely to be 
dominated by contributions  $\sim \langle \bar q G^2 q\rangle$ and $\sim m_s\langle G^3\rangle$
which are not taken into account. Because of this uncertainty, we prefer to stay with the one-resonance fits
in the present case.

\begin{figure}[t]
\centerline{\epsfysize5.7cm\epsffile{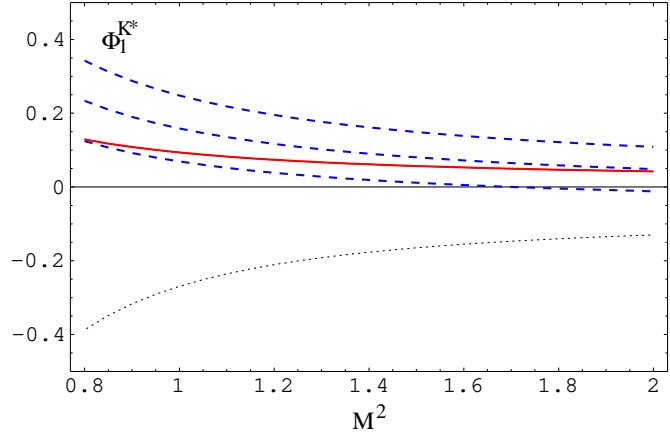}}
\caption[]{\small
 The asymmetry parameter $\phi^{K^*}_1$
obtained from the sum rule (\ref{Patricia}) \cite{BBog} (black dots), and from the sum rules 
(\ref{we2}) (solid red curve) and (\ref{we1}) (dashed blue curves) complemented 
with the identity (\ref{relation1}), as a function of the Borel parameter.
The three dashed curves corresponding to the sum rule in (\ref{we1}) are obtained using the 
values $\delta_5=0$, 0.1 and 0.2 from bottom  to top, respectively.
 }
\label{comparison}
\end{figure}

The results for the asymmetry parameter $\phi^{K^*}_1$
obtained from the sum rule (\ref{Patricia}) on one hand, and from the sum rules 
(\ref{we1}) and (\ref{we2}) with the 
help of the identity  (\ref{relation1}) on the other hand
are shown in Fig.~\ref{comparison} as a function of the Borel parameter.  
In this comparison we discard the last third term ${\cal O}(m^2_s)$ in (\ref{relation1}) because 
the sum rules are written to the ${\cal O}(m_s)$ accuracy only. We see that the both sum rules 
suggested in this work, (\ref{we1}) and (\ref{we2}), support a small and positive value 
of $\phi^{K^*}_1$ which is consistent with the intuition inherited from quark models but 
contradicts the sum rule (\ref{Patricia}). As explained above, this sum rule suffers from 
large cancellations. It is not reliable and has to be discarded.

%
%%%%%%%%%%%%%%%%%%     FIGURE 1          %%%%%%%%%%%%%%%%%%%%%%%%%%%%
%%%%%%%%%%%%%%%%%%%%%%%%%%%%%%%%%%%%%%%%%%%%%%%%%%%%%%%%%%%%%%%%%%%%%%
%

%
%%%%%%%%%%%%%%%%%%     FIGURE 2          %%%%%%%%%%%%%%%%%%%%%%%%%%%%
%%%%%%%%%%%%%%%%%%%%%%%%%%%%%%%%%%%%%%%%%%%%%%%%%%%%%%%%%%%%%%%%%%%%%%
%

\section{Appendix B}

In this Appendix we consider the set of QCD sum rules that are relevant for the K-meson. 
Our aim here is to estimate the ${\mathcal O}(m_s)$ corrections to the result in (\ref{surprise}) and 
present an update for the $SU(3)$ breaking parameter $\delta_5$ for the mixed 
quark-gluon condensate (\ref{delta5}). To this end, following \cite{KKZ87,BD92,AM95},
we consider the correlation functions:
\begin{equation}
 i\int d^4y e^{-iqy}\langle 0|{\rm T}\{\bar u(0) igG_{\mu\nu}{\gamma^\nu}\gamma_5s(0) 
  \bar s(y)\gamma_{\alpha}\gamma_5u(y)\}|0\rangle = 
  g_{\mu\alpha}\Pi_{G,1}^{(a)}(q^2)+ q_\mu q_\alpha \Pi_{G,2}^{(a)}(q^2)
\label{PGaxial}
\end{equation} 
and 
\begin{equation}
 i\int d^4y e^{-iqy}\langle 0|{\rm T}\{\bar u(0) igG_{\mu\nu}{\gamma^\nu}\gamma_5s(0) 
  \bar s(y)i\gamma_5u(y)\}|0\rangle = 
  i q_\mu\Pi_{G}^{(p)}(q^2)
\label{PGpseud}
\end{equation} 
The operator product expansion for the three invariant functions introduced above reads
\begin{eqnarray}
 \Pi_{1,G}^{(a)}(q^2) &=&  -\frac{m_s\alpha_s}{3\pi}
   \left[ \frac53 \langle\bar u u\rangle - \langle\bar s s\rangle\right]\ln\frac{-q^2}{\mu^2}
  - \frac{m_s}{4q^2} \langle \bar u\sigma gG  u\rangle
  + \frac{m_s}{12q^2}\langle\bar s\sigma gG  s \rangle
\nonumber\\&&{}
-\frac{8\pi\alpha_s}{27 q^2}\big[\langle \bar u u\rangle^2 -\langle \bar s s \rangle^2\big],
\nonumber\\
 \Pi_{2,G}^{(a)}(q^2) &=&  +\frac{2m_s\alpha_s}{9\pi q^2}\langle\bar u u\rangle
   \left[\frac13 + \ln\frac{-q^2}{\mu^2}\right]
  -\frac{10m_s\alpha_s}{9\pi q^2}\langle\bar s s\rangle
+ \frac{m_s}{6q^4}\langle\bar s\sigma gG  s \rangle
\nonumber\\&&{}
+\frac{8\pi\alpha_s}{27 q^4}\big[\langle \bar u u\rangle^2 -\langle \bar s s \rangle^2\big],
\nonumber\\
 \Pi_{G}^{(p)}(q^2) &=& 
-\frac{m_s\alpha_s}{48\pi^3} q^2\left[\ln^2\frac{-q^2}{\mu^2}-\ln\frac{-q^2}{\mu^2}\right]
-\frac{\alpha_s}{3\pi} \left[ \langle\bar u u\rangle - \langle\bar s s\rangle\right] \ln\frac{-q^2}{\mu^2}
\nonumber\\&&{}
 -\frac{1}{4q^2}\left[\langle \bar u\sigma gG  u\rangle -\langle\bar s\sigma gG  s \rangle\right]
     +\frac{m_s}{8 q^2}\left \langle\frac{\alpha_s}{\pi}G^2\right\rangle
       \left[ 1 -  \ln\frac{-q^2}{\mu^2}\right]
\nonumber\\&&{}
+\frac{4\pi\alpha_s m_s}{27 q^4}
\left[3 \langle\bar u u\rangle^2 +\langle\bar s s\rangle^2 -9 
     \langle\bar u u\rangle \langle\bar s s\rangle\right].  
\label{PGaxialOPE}
\end{eqnarray}
Our expressions agree with the results of \cite{AM95}. In particular the sign of the perturbative 
contribution to the pseudoscalar correlation function is different from that in \cite{BD92}.
In this calculation we have not included corrections $\sim m_s^2$. Experience of QCD sum rule 
calculations shows that such corrections are tiny and do not influence the results.

Next, we write down the sum rules in which we saturate the phenomenological side by the 
$K$-meson in $\Pi_{G}^{(p)}(q^2)$, $K_1(1270)$ in $\Pi_{1,G}^{(a)}(q^2)$, and both 
$K$ and $K_1(1270)$ in  $\Pi_{2,G}^{(a)}(q^2)$. 
The contribution of the $K_1$ meson is written in terms of the couplings 
\begin{eqnarray}
   \langle 0| \bar u \gamma_\nu\gamma_5  s|K_1(q,\lambda)\rangle &=&  e^{(\lambda)}_\nu f_{K_1}m_{K_1}\,,
\nonumber\\
  \langle 0| \bar u igG_{\nu\mu}\gamma^\mu\gamma_5 s |K_1(q,\lambda)\rangle &=&  
 e^{(\lambda)}_\nu f_{K_1}m^3_{K_1} \kappa_4^{K_1}\,. 
\label{K1}
\end{eqnarray}
Note that the continuum thresholds, in general,  have to be chosen differently for the three cases. 
We obtain the sum rules:
\begin{eqnarray}
\lefteqn{\hspace*{-1.9cm}-f_{K_1}^2 m_{K_1}^4\kappa_4^{K_1} e^{-m_{K_1}^2/M^2} =}
\nonumber\\&=& 
   \frac{m_s\alpha_s}{3\pi} \left[ \frac53 \langle\bar u u\rangle - \langle\bar s s\rangle\right]
    \int_0^{s_{01}}\!\!\! ds\, e^{-s/M^2} 
  +\frac{m_s}{4} \langle \bar u\sigma gG  u\rangle
  - \frac{m_s}{12}\langle\bar s\sigma gG  s \rangle
\nonumber\\&&{}
+\frac{8\pi\alpha_s}{27}\big[\langle \bar u u\rangle^2 -\langle \bar s s \rangle^2\big],
\label{SRK1}
\end{eqnarray}
\begin{eqnarray}
\lefteqn{\hspace*{-0.5cm}
     f_{K}^2 m_{K}^2\kappa_4^{K} e^{-m_{K}^2/M^2} 
    +f_{K_1}^2 m_{K_1}^2\kappa_4^{K_1} e^{-m_{K_1}^2/M^2} = }
\nonumber\\&=& 
   \frac{2m_s\alpha_s}{9\pi}\langle\bar u u\rangle
   \left[- \frac13 + \gamma_E -\ln\frac{M^2}{\mu^2}
+ \int_{s_{02}}^\infty \frac{ds}{s} e^{-s/M^2}\right]
  +\frac{10m_s\alpha_s}{9\pi}\langle\bar s s\rangle
+ \frac{m_s}{6M^2}\langle\bar s\sigma gG  s \rangle
\nonumber\\&&{}
+\frac{8\pi\alpha_s}{27 M^2}\big[\langle \bar u u\rangle^2 -\langle \bar s s \rangle^2\big],
\label{SRK2}
\end{eqnarray}
\begin{eqnarray}
\lefteqn{\frac{f_{K}^2 m_{K}^4}{m_u+m_s}\kappa_4^{K} e^{-m_{K}^2/M^2}=}
\nonumber\\ &=& 
-\frac{m_s\alpha_s}{48\pi^3}
\int_0^{s_{03}}\! ds\, s \, e^{-s/M^2}\left[1-2 \ln \frac{s}{\mu^2}\right]
+\frac{\alpha_s}{3\pi} \left[ \langle\bar u u\rangle - \langle\bar s s\rangle\right]
\int_0^{s_{03}}\!\!\! ds\, e^{-s/M^2}
\nonumber\\&&{}
 +\frac{1}{4}\left[\langle \bar u\sigma gG  u\rangle -\langle\bar s\sigma gG  s \rangle\right]
     +\frac{m_s}{8}\left \langle\frac{\alpha_s}{\pi}G^2\right\rangle
       \left[ - 1 -  \gamma_E + \ln\frac{M^2}{\mu^2}- \int_{s_{03}}^\infty \frac{ds}{s} e^{-s/M^2}\right]
\nonumber\\&&{}
+\frac{4\pi\alpha_s m_s}{27 M^2}
\left[3 \langle\bar u u\rangle^2 +\langle\bar s s\rangle^2 -9 
     \langle\bar u u\rangle \langle\bar s s\rangle\right].  
\label{SRKp}
\end{eqnarray}
One finds by inspection that all three sum rules are dominated by the contributions of the mixed 
quark-gluon condensate. The pseudoscalar sum rule in Eq.~(\ref{SRKp}) is very 
sensitive to the $SU(3)$ breaking in the mixed condensate, whereas  for the other two sum rules
this effect is marginal. One also finds that the sum rules are only very weakly 
sensitive to the values of the continuum thresholds, so we accept $s_{01}=s_{02}=1.8$~GeV$^2$ and
$s_{03}=1.05$~GeV$^2$. We also use the range of Borel parameters $1< M^2 < 2$~GeV$^2$ in all cases. 

\begin{figure}[t]
\centerline{\epsfysize5.2cm\epsffile{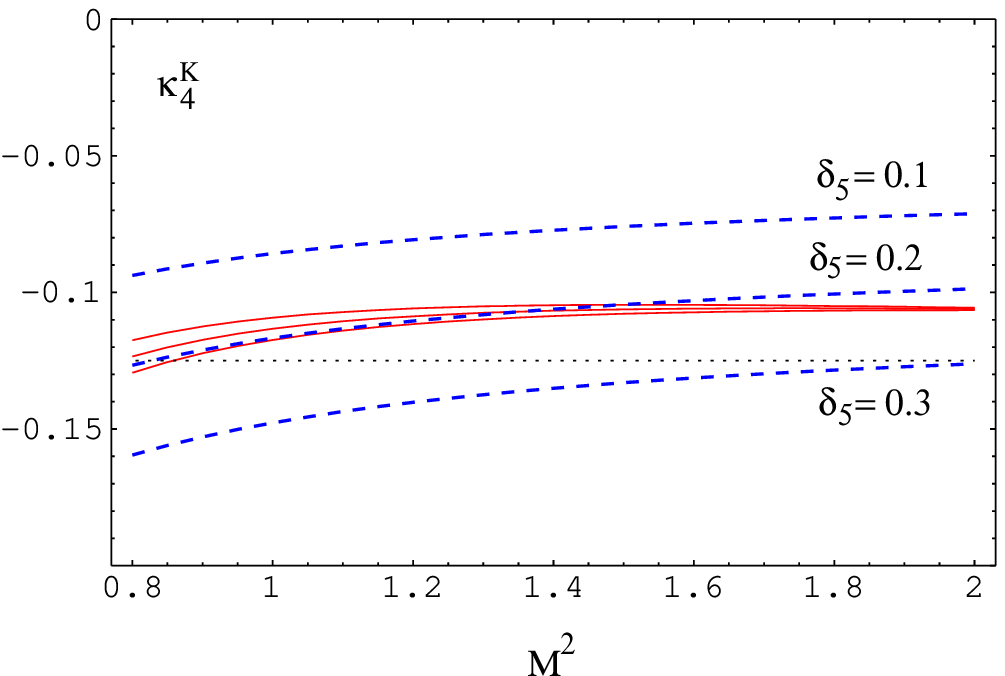}
            \epsfysize5.2cm\epsffile{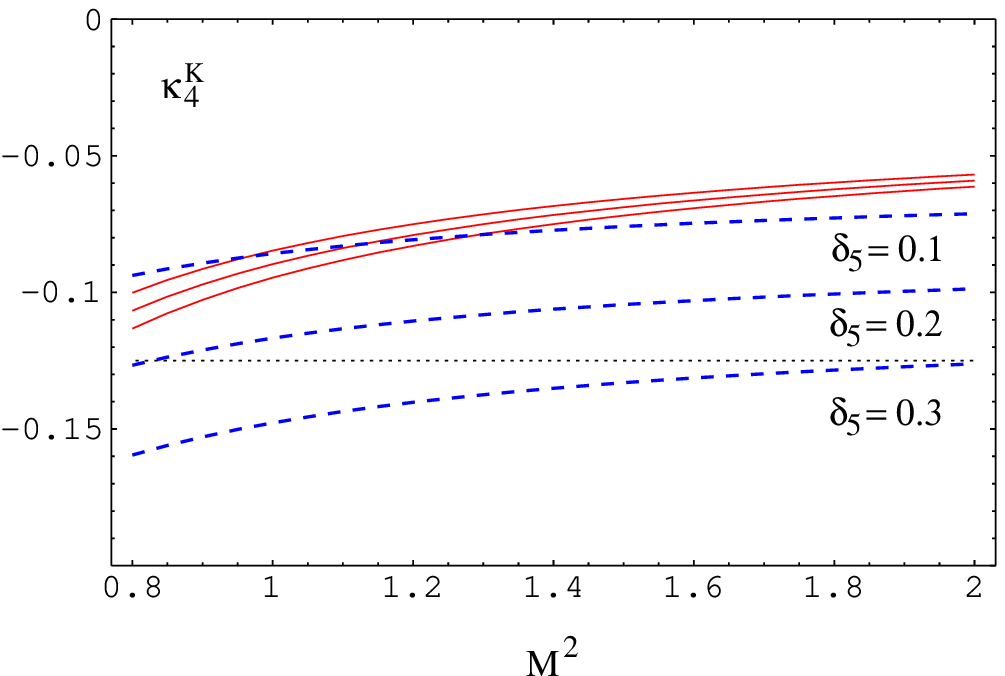}
}
\caption[]{\small
 Left panel: The quark-antiquark-gluon coupling  $\kappa_4^{K}$ from the sum rules 
(\ref{SRK2}) (solid red curves) and  (\ref{SRKp}) (dashed blue curves)
for three different values of $\delta_5$ as a function of the Borel parameter.
The exact value $\kappa_4^{K} =-0.125 $ in the $SU(3)$ symmetry limit 
(\ref{surprise}) is shown by dots for comparison.
Right panel: The same, but neglecting  the $K_1$-meson contribution in the 
sum rule (\ref{SRK2}).
 }
\label{kappaK}
\end{figure}

Next, we adopt the following procedure. 
On one hand, we use the pseudoscalar sum rule (\ref{SRKp}) to calculate the coupling $\kappa_4^K$ as
a function of the Borel parameter $M^2$ for three different values $\delta_5 =0.1, 0.2, 0.3$. 
On the other hand, we use the sum rule in (\ref{SRK1}) to get an estimate for the relevant product of 
couplings of the $K_1$ meson:
\begin{equation}
                f^2_{K_1}\kappa_4^{K_1} = \left(0.36^{+0.22}_{-0.08}\right)\cdot 10^{-3}~\mbox{\rm GeV}^3
\end{equation}
and then substitute this number in the sum rule (\ref{SRK2}) which we, again, evaluate for 
three different values of $\delta_5$. The results are shown in Fig.~\ref{kappaK}, left panel.
{}For illustration, we also show (right panel) the results obtained by neglecting the $K_1$ meson contribution 
to the axial-vector sum rule (\ref{SRK2}). Note that this contribution has opposite sign $\kappa_4^{K_1} > 0$
compared to that of the $K$-meson: $\kappa_4^{K} < 0$.

%
%%%%%%%%%%%%%%%%%%     FIGURE 3          %%%%%%%%%%%%%%%%%%%%%%%%%%%%
%%%%%%%%%%%%%%%%%%%%%%%%%%%%%%%%%%%%%%%%%%%%%%%%%%%%%%%%%%%%%%%%%%%%%%
%

{}From  Fig.~\ref{kappaK} we conclude that the axial vector and the pseudoscalar sum rules prove to be 
in a very  good agreement with each other, provided one uses the standard values of the sum rule parameters
and in particular $\delta_5=0.2$. The choice of 
$m_0^2 = \langle \bar q \sigma gG q\rangle/\langle \bar q q\rangle$ and the strange quark mass 
within the accepted range does not have significant influence on the preferred value of $\delta_5$.
On the other hand, the value of $\delta_5$ turns out to be correlated with   
$\delta_3 = 1- \langle \bar s s\rangle /\langle \bar u u\rangle$. Taking $\delta_3 = 0, 0.2,0.4$ 
one obtains the best agreement between the sum rules for $\delta_5 = 0.3, 0.2, 0.1$, respectively.   
These values are somewhat larger than the estimate $\delta_5 \simeq 0.05$ obtained in Appendix A 
from the comparison of chirality-breaking vector-vector and vector-tensor correlation function,
and we believe that the present analysis is more reliable. The reason for this is that in the 
present case the sum rules appear to be more stable and, more importantly, 
we can make two-resonance fits taking advantage to determine the $K_1$ 
contributions from a separate sum rule (\ref{SRKp}) rather than relying on the requirement of stability. 
Note that the one-resonance fit to the axial vector sum rule, which is similar to our 
fits in Appendix A, also tends to support a smaller value $\delta_5\sim 0.1$, see the right panel
in Fig.~\ref{kappaK}.  
We give the ``weighted'' average of the estimates from the both channels as our final 
result in Eq.~(\ref{d5final}).

The coupling $\kappa_4^K$ calculated from the sum rules using experimental values for $f_K$ and $m_K$ 
turns out to be somewhat smaller (in absolute value) than the corresponding result $\kappa_4^{K} =-0.125$
in the chiral limit, see  Fig.~\ref{kappaK}, and is mostly affected by the value of $m_0^2$. 
Our final result for $\kappa_4^K$ is given in Eq.~(\ref{kaSR}), see text.

\end{document}